\documentclass[sigconf]{acmart}
\AtBeginDocument{%
  \providecommand\BibTeX{{%
    \normalfont B\kern-0.5em{\scshape i\kern-0.25em b}\kern-0.8em\TeX}}}

\copyrightyear{2023}
\acmYear{2023}
\setcopyright{acmlicensed}\acmConference[WebSci '23]{15th ACM Web Science Conference 2023}{April 30-May 1, 2023}{Evanston, TX, USA}
\acmBooktitle{15th ACM Web Science Conference 2023 (WebSci '23), April 30-May 1, 2023, Evanston, TX, USA}
\acmPrice{15.00}
\acmDOI{10.1145/3578503.3583629}
\acmISBN{979-8-4007-0089-7/23/04}

\usepackage{multirow}
\usepackage{subfig}

\begin{document}

%\title{Employment gender gaps: sub-national insights in Italy using LinkedIn Advertising data}

\title{Monitoring Gender Gaps via LinkedIn Advertising Estimates: the case study of Italy}
%\title{LinkedIn ads data for a sub-national insight into employment gender gaps in Italy}


\author{Margherita Bertè}
\affiliation{
  \institution{ISI Foundation}
  \city{Turin}
  \country{Italy}}
\email{margherita.berte@isi.it}
\orcid{0000-0003-4372-9734}

\author{Kyriaki Kalimeri}
\affiliation{
  \institution{ISI Foundation}
  \city{Turin}
  \country{Italy}}
\email{kyriaki.kalimeri@isi.it}
\orcid{0000-0001-8068-5916}

\author{Daniela Paolotti}
\affiliation{
  \institution{ISI Foundation}
  \city{Turin}
  \country{Italy}}
\email{daniela.paolotti@gmail.com}
\orcid{0000-0003-1356-3470}

\renewcommand{\shortauthors}{Bertè M. et al.}

\begin{abstract}

Women remain underrepresented in the labour market. Although significant advancements are being made to increase female participation in the workforce, the gender gap is still far from being bridged.
We contribute to the growing literature on gender inequalities in the labour market, evaluating the potential of the LinkedIn estimates to monitor the evolution of the gender gaps sustainably, complementing the official data sources. 
In particular, assessing the labour market patterns at a subnational level in Italy.
Our findings show that the LinkedIn estimates accurately capture the gender disparities in Italy regarding sociodemographic attributes such as gender, age, geographic location, seniority, and industry category.
At the same time, we assess data biases such as the digitalisation gap, which impacts the representativity of the workforce in an imbalanced manner, confirming that women are under-represented in Southern Italy.
Additionally to confirming the gender disparities to the official census, LinkedIn estimates are a valuable tool to provide dynamic insights; we showed an immigration flow of highly skilled women, predominantly from the South. 
Digital surveillance of gender inequalities with detailed and timely data is particularly significant to enable policymakers to tailor impactful campaigns.
\end{abstract}




\received{30 November 2022}
\received[revised]{31 January 2023}
\received[accepted]{28 February 2023}

\maketitle

\section{Introduction}\label{sec:Introduction}

Despite being a fierce debate for decades, gender discrimination in the workplace remains an active issue. In fact, 
``Gender Equality'' and ``Decent work and economic growth'' are among the top Sustainable Development Goals (SDGs)~\cite{UN2020} of the United Nations agenda for 2030.
Gender inequalities occur across various domains including education~\cite{Ellis2016}, life expectancy~\cite{Pinho-Gomes2022} , personality and interests~\cite{Eagly2019}, family life~\cite{Stojmenovska2020}, and careers~\cite{Cheryan2013, Benson2021, Azmat2021}. 
Hence, bridging the gender gap in the workplace is far from a trivial topic, as it entails significant welfare and cultural changes.
Gender inequality in the workplace may take various forms ranging from unequal pay and promotion disparity~\cite{Benson2021, Azmat2021} to incidents of sexual harassment~\cite{Bosco2021, Harnois2018, Folke2020, Shaw2018}.
Those often occur in a nuanced way rendering the phenomenon difficult to 
thoroughly quantify.

Each year, the World Economic Forum publishes the Global Gender Gap Report~\cite{WEF2022} containing the status and the steps forward to close the gender gap in 146 countries. Some problems emphasized are the need for more women in jobs related to STEM disciplines (science, technology, engineering and mathematics), the absence of leadership positions and the need for full and effective female participation in the labour force.
The gender-disaggregated data around these inequalities are often lacking, which hinders welfare policies due to the incomplete view of the problem.  
Recently, scientists employed data from social media (SM) and crowdsourcing platforms to provide a complementary view of society and overcome the known time and cost limitations of official surveys~\cite{Kashyap2020, Kashyap2021, Haranko2018, verkroost2020tracking, Tamime2022}. 
The potentials of SM advertising platform estimates have been validated on a wide range of topics, including the assessment of wealth ~\cite{Fatehkia2020}, the rural-urban divide~\cite{Rama2020}, and migration flows~\cite{zagheni2017leveraging}. Particularly for sparse populations, social media offer a valid alternative to track important socioeconomic statistics~\cite{Mejova2018,Fatehkia2019,Rama2020, Fatehkia2020}.

In this study, we examine the benefits and validity of the LinkedIn Advertising (ads) platform as a potential data source for the labour market digital ``census''.
LinkedIn is the world's largest social networking platform targeted at professionals with
a user base of over 900 million spanning over 200 countries (membership members)~\cite{LinkedInstat2021}.
Although the advertising platform was initially designed to estimate audience reach, scientists have leveraged its potential to study gender disparities~\cite{WEF2022,verkroost2020tracking,Kashyap2021}. At the same time,  establishing a coherent metric to assess the gender gap systematically is a field of active research~\cite{Cascella2022}.

To face the employment gender disparities using LinkedIn data from a sub-national perspective, we focused on Italy as a case study.
Its geographical divide is an unsolved theme with a broad literature.
The challenge is understanding whether gender gaps in the workplace, known from traditional survey data ~\cite{DiMartino2020, Cascella2022, diBella2019, Greselin2020, Campa2011}, are reflected in the data obtained through the LinkedIn social media platform.
Italy is the third largest economy in the European Union ~\cite{EUROSTAT2021}, and although in the last years, significant progress has been made in reducing gender inequalities in the labour market, they persist in being tightly woven into the social fabric. 
Here, we examine those from a sub-national point of view, relating LinkedIn's estimates to measures provided by traditional data obtained through the Italian national institute of statistics (ISTAT)~\cite{ISTAT2022} and the European Statistical Office (EUROSTAT)~\cite{EUROSTAT2022}.
Local perspective is particularly significant in Italy since, by Constitution, regional administrations can act directly to mitigate the problem.
Finally, the need for a wide Italian presence on LinkedIn is met: it is the third European country by the number of members, with about 17 million users~\cite{LinkedInstat2021}.

We aim to address the following core research questions:\\
    \textbf{RQ1} How reliable are LinkedIn advertising audience estimates, especially concerning a country’s labour force and the official
demographic figures?\\
     \textbf{RQ2} How can we enrich the current view of the gender gap in Italy as seen through LinkedIn?\\
    \textbf{RQ3} Can we predict the employment gender gap leveraging LinkedIn estimates and sociodemographic
data?

To answer the aforementioned questions, we estimated the Gender Gap Index (GGI)~\cite{verkroost2020tracking} on the numeric estimates of the potential audiences obtained via the platform's API. 
We show that LinkedIn's population estimates correspond to the official Italian labour force census. 
The gender distribution in each economic sector on LinkedIn positively correlates with the official data. Nonetheless, we observe that most sectors are under-represented, except those related to Technology, which is coherent with the highly skilled workforce the platform addresses to~\cite{LinkedInskil2021}.

LinkedIn has its inherent population biases with gender and age ranges not uniformly represented~\cite{Kashyap2021}. However, the data obtained by the platform are representative of the labour force in general, with the increased gender gap observed in the Southern part of Italy~\cite{Cascella2022} and the more senior roles having higher male estimates in leadership positions~\cite{WEF2022}.
This aligns with broader known sociodemographic inequalities in Italy~\cite{Buzzacchi2021, DiMartino2020}.
The digitalisation rate reported by ISTAT impacts the representativeness of the workforce in an imbalanced manner, with women needing to be more represented in Southern Italy.
This is a crucial point to consider when leveraging this data source to assess the labour force gender gap in countries with a low digitalisation rate.
Finally, we noticed that highly skilled employees (graduated or doctorate level) are more likely to move to other countries, undermining the development of the labour force, particularly in the most vulnerable areas (e.g. the South). The LinkedIn audience reflects this: the platform is more gender-balanced in the regions with more high-skilled female immigrants from abroad (e.g. North-Center).
We contribute to the current literature by showing the importance of the LinkedIn advertising Platform in accurate monitoring 
the labour patterns within countries, focusing on Italy as a case study.

\section{Related work}\label{sec:related_work}
% traditional gender gap assessment

Traditionally, the gender divide is monitored with data provided by Census.
The Global Gender Gap Index (GGGI) was theorised in 2006~\cite{WEF2006} and since then is employed to compare the different aspects of the gender gap (economy, health, education, politics) worldwide.
In 2013, the Gender Equality Index was computed for the European Union, commissioned by the European Institute for Gender Equality (EIGE)~\cite{EIGE2010}.
Those indices allow us to keep track of the gendered disparity and compare data worldwide and even across years; however, the sub-national disparities were not assessed.

%\textbf{Digital Gender Gaps.} 
\paragraph{Digital Gender Gaps.}Over the last few years, scientists have relied increasingly on digital data estimates and SM advertising platform estimates to address complex demographic and social research questions.
Early on, Garcia et al.~\cite{Garcia2018} suggested a global measure for gender disparity leveraging on Facebook ad estimates for 217 countries, while more recently, Fatehkia et al.~\cite{Fatehkia2019} investigated the digital gender gap for 193 countries, confirming known trends obtained via traditional sources. 
Combining Facebook with Google advertising data, Kashyap et al.~\cite{Kashyap2020} assessed the worldwide digital divide. 
The high resolution and richness of SM advertising estimates allow for sub-national focus~\cite{Mejova2018}, but also for evaluating gender inequalities concerning specific interests such as the interest in the STEM disciplines
%stem studies
\cite{Tamime2022,Vieira2021}.

\paragraph{Labour Market Gender Gaps.} The LinkedIn advertising platform offers critical demographic estimates facilitating the study of various topics relating to labour dynamics~\cite{verkroost2020tracking, Kashyap2021}.
LinkedIn estimates were informative to understand the variations of gender gaps in IT industries both globally~\cite{verkroost2020tracking, Kashyap2021} and sub-nationally~\cite{Haranko2018}. 
The LinkedIn Gender Gap Index (GGI) was initially proposed in ~\cite{Kashyap2021, verkroost2020tracking} as the ratio between the estimated number of women with specific attributes over the estimated number of men with the same attribute, consisting the equivalent of GGGI for LinkedIn ad estimates. Since then, varied metrics have been proposed, but the GGGI was shown to be the most accurate to measure the gender gap~\cite{Cascella2022}.

Although this study is close to the approach and methods presented in Verkroost et al.~\cite{verkroost2020tracking}, and Kashyap et al.~\cite{Kashyap2021}, the metric we employ to assess the labour gender gap slightly differs from the ones proposed in the present literature~\cite{verkroost2020tracking, Kashyap2021,WEF2022}.
In particular, to capture the interplay of the local dynamics in Italy, we dive into the Italian regions' labour market data,  normalise the estimates according to the official population census data, and we do not a priori assume the female gender to be under-represented. 
%As so, we avoid the binary vision of females as under-represented/other.
The GGGI benchmarks the current state and evolution of women's situation in four key dimensions (Economic Participation and Opportunity, Educational Attainment, Health and Survival, and Political Empowerment); here we explore disparities for both genders.

\begin{figure*}[!ht]
    \includegraphics[width=0.8\textwidth]{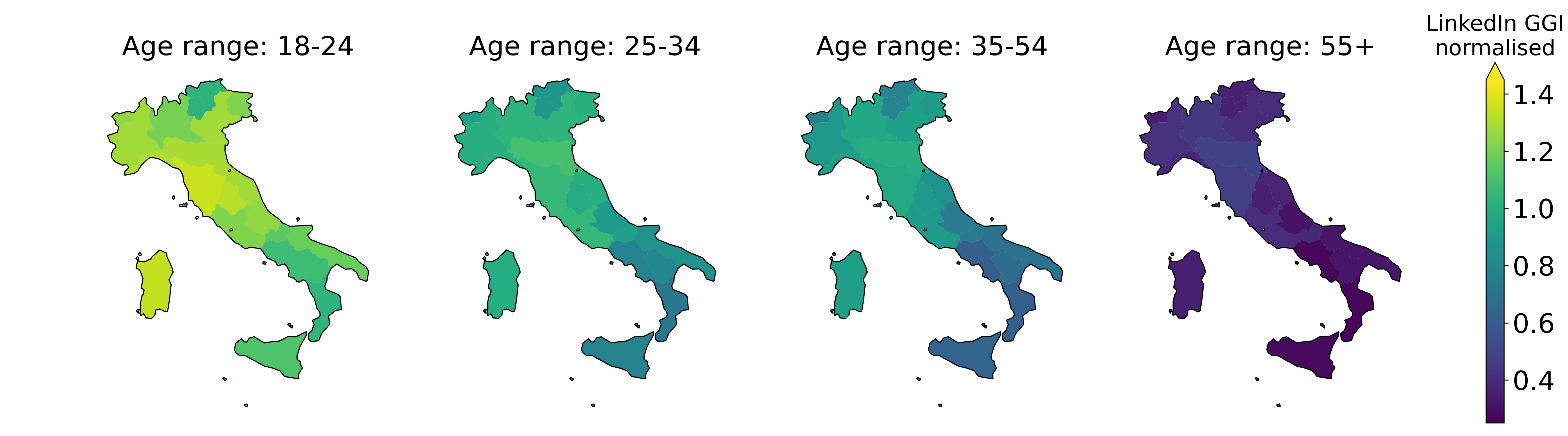}
    \caption[\textsf{LinkedIn GGI normalised} with exclusion query by age range and regions]{Representation of \textsf{LinkedIn GGI normalised} in the Italian regions with exclusion query.}
    \label{fig:GGI_LinkedIn_geopandas}
\end{figure*}

\begin{table*}[hbtp]
        \caption[Italian estimates: LinkedIn vs Census]{Number of people by age range estimated by LinkedIn ads in September 2022 and in ISTAT as reported by 2021 data. We also report the percentages of LinkedIn's estimated audience over the census population numbers reported by ISTAT for each age range.}
        \label{tab:italy_age_linkedin_census}
        \small
        \begin{tabular}{lllllll} 
         \toprule
        Age & \textsf{F LinkedIn} & \textsf{M LinkedIn}  & \textsf{F Census} & \textsf{M Census} & \textsf{F LinkedIn} / \textsf{F Census} & \textsf{M LinkedIn} / \textsf{M Census}\\
         \midrule
        18-24  &  1.300.000 & 1.200.000 & 1.963.785 & 2.136.690 & 66,20\%  & 56,16\% \\
        25-34  &  4.600.000 & 4.700.000 & 3.048.664 & 3.195.963 & 150,89\% & 147,06\% \\
        35-54  &  1.700.000 & 2.100.000 & 8.371.114 & 8.279.348 & 20,31\% & 25,36\% \\
        55+  &   180.000  & 380.000   & 12.371.753 & 10.396.360 & 1,45\% & 3,66\% \\
         \bottomrule
        \end{tabular}
\end{table*}

\section{Data Collection}

For this study, we obtained data from the European and Italian official data statistics offices, i.e. EUROSTAT and ISTAT, respectively, and from the LinkedIn ads platform~\cite{LinkedInads2016}.

 \paragraph{Official Census and Survey Data.} Italy can be divided into five main geographical zones. ISTAT computes the rate of regular Internet users through a survey to measure how many people have daily access to the Internet, regardless of the device. Based on this yearly report, 
there is a uniform digital gender gap throughout Italy (on average, six percentage points difference between the rate of regular male Internet users from the female rate); in the North, the average rate of regular Internet users is 78\% for men and 72\% for women; in the South, the average is 71\% for men and 65\% for women.

\begin{itemize}
    \item North-East: Emilia-Romagna, Friuli-Venezia Giulia, Trentino Alto Adige, and Veneto
    \item North-West: Liguria, Lombardia, Piemonte, and Valle d'Aosta
    \item Center: Lazio, Marche, Toscana, and Umbria
    \item South: Abruzzo, Campania, Molise, Puglia, Basilicata, and Calabria
    \item Islands: Sardegna and Sicilia
\end{itemize}

These five zones are merged into two main groups: North-Center (North-East, North-West, Center) and South (South and Islands), composed of regions quite similar in socioeconomic status.
The locations for the data collection are chosen taking the European official territorial units for statistics (NUTS from the French acronym) of level 2 for 2021 provided by EUROSTAT~\cite{NUTS2018}.

\paragraph{LinkedIn ad Audience Estimates.}
Focusing on the Italian context, we collected aggregated counts of LinkedIn users, querying the ad campaign manager via the official application programming interface (API)\footnote{We were based on the open source code by Lucio Melito \url{https://worldbank.github.io/connectivity\_mapping/intro.html}}. 
Audiences are targeted based on geographic location, demographic criteria such as gender or age group, and job criteria such as company industry and job seniority.
Here, we target the locations to a sub-national level to capture the local nuances and trends that are otherwise difficult to obtain. 
Overall, we gathered data for 20 regions in Italy from July to November 2022.
In detail, we queried for the following characteristics (attributes).

\begin{itemize}
    \item \textbf{Location.} According to LinkedIn official documentation~\cite{LinkedIn}, this attribute can be based on the location a member has included in their profile or their IP address. We collected the Italian data referring to the NUTS2, i.e. the basic regions for applying regional policies. For Italy, they are 20 regions.
    \item \textbf{Gender.}  On the LinkedIn ads platform, gender is binary: Male, Female. In our study, hence we follow this binary approach. Among the know limitations of our work, we acknowledge the  binary choice of gender and the overarching assumption that all LinkedIn users are active in the labour market.
    \item \textbf{Age range.} Age is provided in the following ranges 18-24, 25-34, 35-54, 55+. The Age range a member belongs to, is inferred on their first graduation date, but also the \emph{Years of Experience} can be used as a proxy as to what is reported by the official  documentation~\cite{LinkedIn}. 
    \item \textbf{Job seniority.} It ``describes the rank and influence of a member's current role in their organization'' (as stated in~\cite{LinkedIn}). We target all seniority levels Unpaid, Training, Entry, Senior, Manager, Director, VP, CxO, Partner, and Owner.
    \item \textbf{Company industry.} The economic sector in which the employing company belongs to~\footnote{The list of the 20 main company industries chosen can be found in Table~\ref{tab:map_industry_NACE} of the Appendix~\ref{sec:appendix}.}. 

\end{itemize}

Table~\ref{tab:italy_age_linkedin_census} summarises the data collection performed per age group and gender compared to the official population census (ISTAT). Further, we gathered the estimates by gender and age range per region.
Figure~\ref{fig:GGI_LinkedIn_geopandas} depicts the estimates obtained per Italian region.

\begin{figure*}[h]
    \includegraphics[width=1\textwidth]{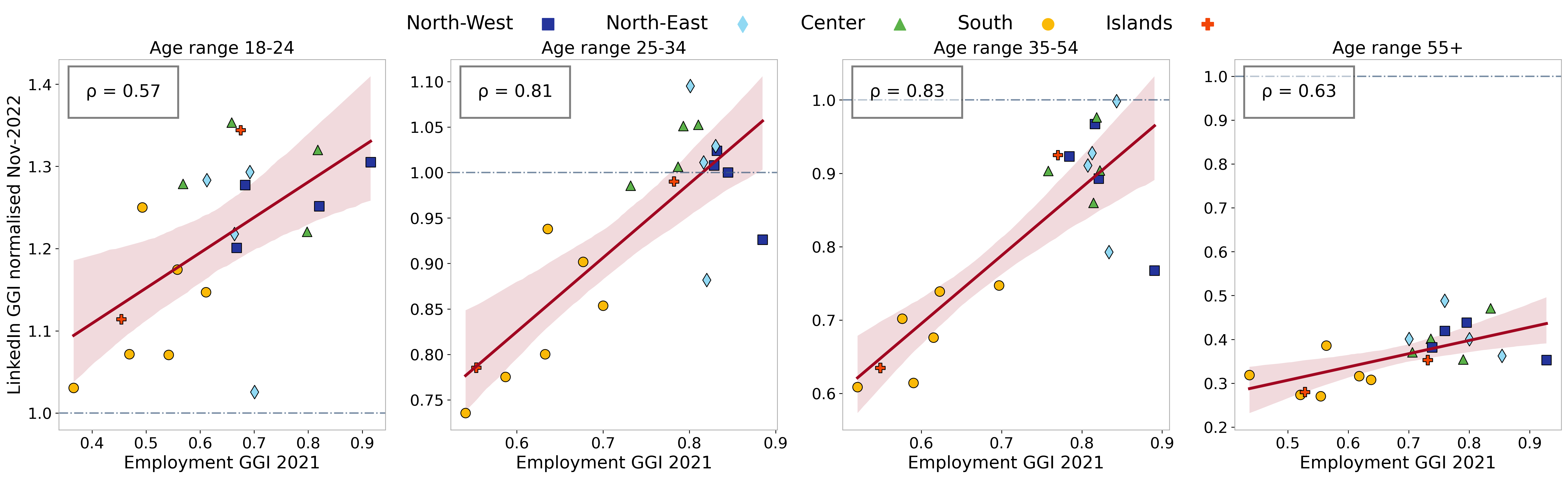}
    \caption[\textsf{LinkedIn GGI normalised} vs ISTAT employment data by regions]{\textsf{LinkedIn GGI normalised} regional estimates (y-axis) versus \textsf{Employment GGI} (x-axis) by age range. Data will be collected for Italian regions in September 2022. The dark red line shows a regression line with 95\% confidence; the light blue line is the equality line. Values above the dashed line (y-axis) indicate an over-representation of women. }
    \label{fig:LinkedIn_ISTAT_empl}
\end{figure*} 

\subsection{Methods}

\paragraph{Data Preparation.}
Overall, for all the locations (20 regions in Italy), we queried by gender and age to obtain a ``LinkedIn Census''.  
At national level, keeping the location fixed (Italy), we targeted the audiences by gender, age-range and job seniority to get seniority data, while to collect estimates per economic sector, we queried by gender, age-range and company industry.
We obtained estimates for all queries (19 regions) but one; the number of women on LinkedIn older than 55 years old located in Valle d'Aosta (the smallest Italian region). In that case,  the estimated number of members was less than 300 users, and hence, the API returns 0. 
To overcome this issue, we employed the ``query exclusion'' method proposed by Rama et al.~\cite{Rama2020}. We obtained an estimate of over 200 members, still sufficient to ensure privacy.
To apply the ``query exclusion'' approach, we chose several ``reference cites'' opting for large cities with a sufficient amount of users for the majority of the categories we are interested in. 
We decided to refrain from collecting data at the provincial level to avoid privacy concerns. For the same reason for queries with more specific targeted attributes (besides location, gender, and age range), we opted to obtain the data at a national level to avoid possible identification of users.

\paragraph{Gender Gap Metrics.}
We measured the gender divide in the LinkedIn community adapting the Global Gender Gap Index (GGGI)~\cite{WEF2022} proposed by Verkroost et al.~\cite{verkroost2020tracking}.
First, we computed the gender gap index of the LinkedIn population normalising ads estimates by ISTAT population data:
\begin{equation}
       \textsf{LinkedIn GGI normalised} = \frac{\textsf{F LinkedIn}}{\textsf{M LinkedIn}} \cdot \frac{\textsf{M Census}}{\textsf{F Census}} 
       \label{eq:Link-GGI-Census}
\end{equation}
\noindent where, $\textsf{F LinkedIn}$ and $\textsf{M LinkedIn}$ are the estimated number of women and men, respectively, in LinkedIn, while $\textsf{F Census}$ and  $\textsf{M Census}$ are the estimated number of women and men, respectively reported by ISTAT.

\noindent For each attribute, we adapt the above equation normalising by the LinkedIn population,
\begin{equation}
        \textsf{LinkedIn GGI (Attribute)} = \frac{\textsf{F LinkedIn (Attr)}}{\textsf{M LinkedIn (Attr)}} \cdot \frac{\textsf{M LinkedIn}}{\textsf{F LinkedIn}}
        \label{eq:Link-attr}
\end{equation}

\noindent where:
\textsf{F LinkedIn (Attr)} refers to the estimated number of women in LinkedIn that have the specific attribute, and \textsf{M LinkedIn (Attr)} to the number of men, respectively.
Contrary to the mainstream approach~\cite{WEF2022}, where an over-estimation of women is not considered (GGI index greater than one), we opted for the whole range of values, also evidencing when men are the minority group.
To measure the gender gap in employment status and economic section we defined the following ratios:

\begin{equation}
    \textsf{Employment GGI} = \frac{\textsf{\% F Working}}{\textsf{\% M Working}}
    \label{eq:empl-GGI}
\end{equation}
\noindent
where \textsf{\% F Working} (resp. \textsf{\% M Working}) is the percentage of working women (resp. men). 
Similarly, for each economic sector, we estimated the index as the gender ratio between the EUROSTAT percentage of employed women and men in each Nomenclature of Economic Activities (NACE):
\begin{equation}
    \textsf{NACE employment GGI} = \frac{\textsf{\% F NACE Working}}{\textsf{\% M NACE Working}}
    \label{eq:empl-NACE}
\end{equation}
\noindent
where \textsf{\% F NACE Working} (resp. \textsf{\% M NACE Working}) indicates the percentage of women (resp. men) working in a particular economic sector.
All the indices introduced in this section were computed by age range and location to unveil discrepancies that may occur based on those factors.

\paragraph{Prediction of Gender Gap.}
\label{par:model_design}
Finally, we assessed the digital estimates' prediction capabilities of the gender gap. 
We built a multilinear regression model that leverages LinkedIn estimates and sociodemographic data to predict the gender gap in the workforce (Eq.~\ref{eq:empl-GGI}). We evaluated the model performance employing the \textsf{Mean Absolute Error} ($MAE$) and the \textsf{$R^2$-adjusted} ($R^2_{adj}$) accounting for the number of predictors. To reduce the dimensionality, we implemented a step-wise feature selection. 
The set of features considered included the age range and several socio-economic regional indicators, such as the percentage of Gross domestic product (GDP) per capita in Purchasing Power Standards (PPS) expressed to the European Union average (equal 100), the gender ratio of digitalization level, the gender ratio of the percentage of young NEET (Not [engaged] in Education, Employment or Training) and a welfare policy marker: the percentage of children under two years old going to the kindergarten.
We applied also 5-fold cross validation method to check overfitting and the Isolation Forest Algorithm (with contamination 0.05) to detect outliers before the model training.

\section{Results \& Discussion}

\subsection{Socioeconomic \& Demographic Representativity.}
This work explores an increasingly popular data source in digital Demography, the LinkedIn Advertising estimates.
These data provide cost-effective insights into gender inequalities in the labour market in a detailed and timely manner, essential to crafting impactful policies, especially when the official data are sparse or hard to obtain. 
Our findings confirm the reliability and potential of the LinkedIn advertising audience estimates to the labour force dynamics recorded from the official sources.

\begin{figure}[htpb]
    \includegraphics[width=0.45\textwidth]{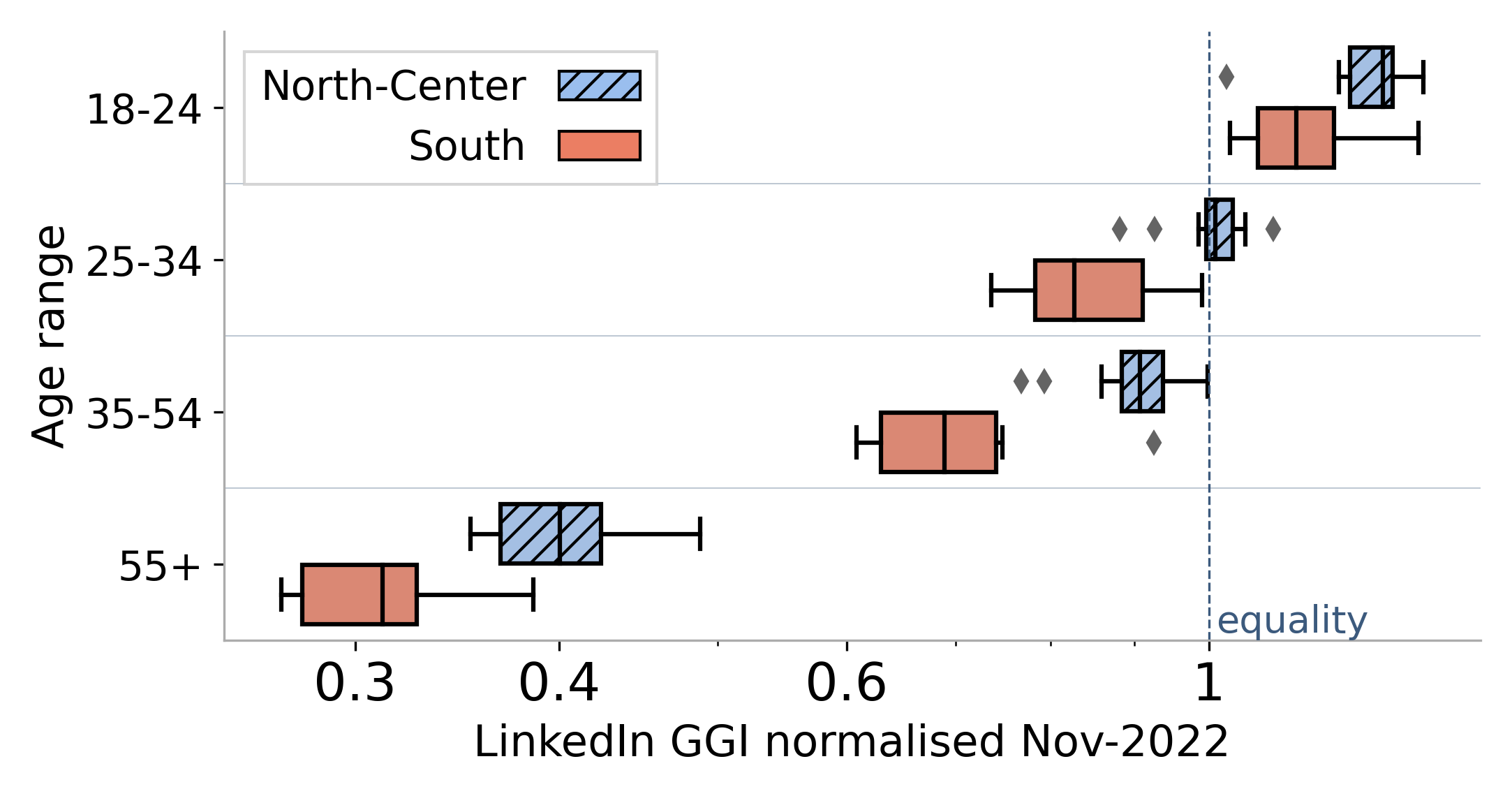}
    \caption{Distribution of the \textsf{LinkedIn GGI normalised} measure across different age ranges (y-axis in log scale) aggregated by zones of Italy.}
    \label{fig:GGI_LinkedIn_age_range}
\end{figure}

\paragraph{Demographic Representativity.} For a digital data source to be reliably employed in policy and decision-making, assessing its demographic representativity is fundamental. Comparing the \textsf{LinkedIn GGI normalised} index (Eq.~\ref{eq:Link-GGI-Census}) with the \textsf{Employment GGI} index (Eq.~\ref{eq:empl-GGI}) per gender and age range (aligning the age groups are reported in Table~\ref{tab:map_age_link_ISTAT}), we noticed a strong positive correlation for all the age groups indicating that the obtained estimates were demographically representative of the Italian workforce (Table~\ref{tab:corr_italy_linkedin_census}, row 2).

Figure~\ref{fig:LinkedIn_ISTAT_empl} depicts the relationship of the gender gap observed in LinkedIn to the official ISTAT employment data. Women are vastly over-represented in the younger age group (\textsf{LinkedIn GGI normalised} above the equity line), while they are under-represented in the elder age group (\textsf{LinkedIn GGI normalised} under the equity line). For the ISTAT data, none of the age ranges reaches the equity line (value one on the x-axis). The percentages of women working are lower than those of men.
In Figure~\ref{fig:LinkedIn_ISTAT_empl}, we also noticed that the Italian regional divide is captured: the Southern regions lack a workforce with respect to the Northern ones (southern regions are grouped in the lower part of the regression line). 
This aligns with the general employment trends in Italy~\cite{Cascella2022, ISTAT2020}.

Focusing on the intrinsic biases of the medium, we observed more male than female users on LinkedIn (Figure~\ref{fig:GGI_LinkedIn_age_range}) for all the age ranges from 25 years old and elder, opposite to what happens in other social media as META or Instagram~\cite{Ribeiro2020}.
From a geographical point of view, \textsf{LinkedIn GGI normalised}'s distribution varies across the country. On average, in the South, we have less gender parity, especially as the audience gets older.
Consistently with the findings of Kashyap et al.~\cite{Kashyap2021}, the most accurate age range representation is observed for the group of 25 to 34 years old, whereas the category of the users being more than 55 years old is strongly limited, for women in particular. In Table~\ref{tab:italy_age_linkedin_census}, the age range 25-34 is over-represented in LinkedIn to ISTAT data in Italy. The same behaviour was observed for the Facebook ad estimates~\cite{Ribeiro2020}. This can be partly attributed to the fact that people may be temporarily located in Italy and hence not recorded by the official Census. Fake accounts may also be part of the equation~\cite{CNN2022}.
Overall, the LinkedIn estimates are shown to capture the employment frame in Italy reliably.

\begin{figure*}[!ht]
    \includegraphics[width=0.95\textwidth]{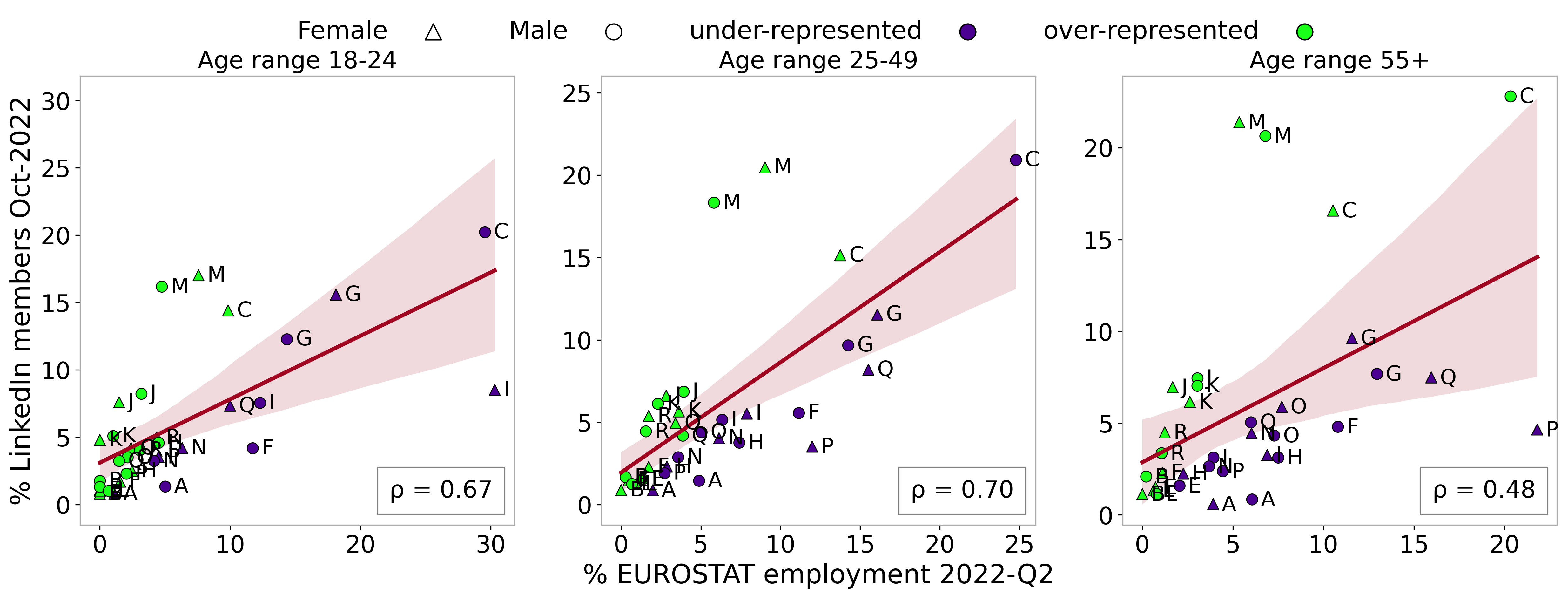}%corr_IT_empl_NACE_merged
    \caption{LinkedIn percentage estimates (y-axis) versus employment percentage (x-axis) by age range, gender, and NACE. The over-represented economic sectors are highlighted. Data were collected for Italy in October 2022. The dark red line shows a regression line with 95\% confidence and the labels refer to the EUROSTAT NACE codes as reported in Table~\ref{tab:map_industry_NACE}, such as:
    \small{
    A: Agriculture,
    C: Manufacturing, 
    F: Construction,
    G: Sales,
    H: Transportation,
    I: Accommodation,
    J: IT and Media,
    K: Financial Services,
    M: Professional Service,
    P: Education,
    Q: Health,
    R: Entertainment}}
    \label{fig:LinkedIn_EU_nace}
\end{figure*}

\begin{figure}[h]
    \includegraphics[width=0.48\textwidth]{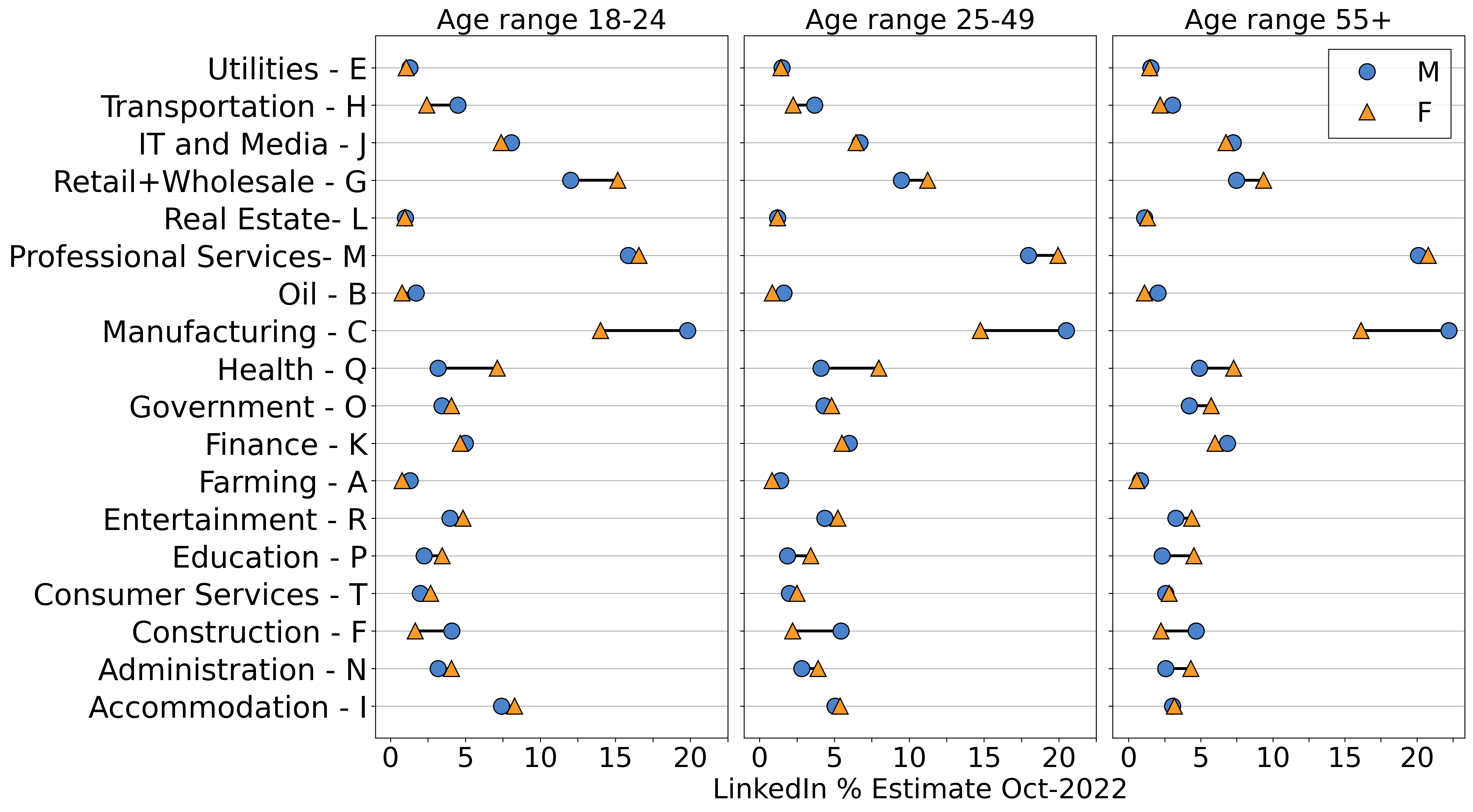}
    \caption{Comparison of LinkedIn percentage estimates by age range, gender, NACE.}    \label{fig:LinkedIn_EU_nace_gender_comparison}
\end{figure}

\paragraph{Company Industry Representativity.} 
To assess possible biases within the economic sectors, we aggregated the LinkedIn users by gender, age, and per each economic sector\footnote{Since age ranges and economic sectors do not correspond perfectly between EUROSTAT and LinkedIn, we grouped them as reported in Table~\ref{tab:map_age_euro_link} and ~\ref{tab:map_industry_NACE}.}. 
For each group, we estimated the percentages of employed individuals over the total number of employees for LinkedIn and EUROSTAT.
Correlating these percentages, we found positive relationships reported in Table~\ref{tab:corr_italy_linkedin_census}.

Figure~\ref{fig:LinkedIn_EU_nace} depicts the relationship between the two data sources, LinkedIn and EUROSTAT.
We notice that on LinkedIn, several economic sectors are under-represented (e.g.  Farming-A, Construction-F, Administration-N, Transportation-H),  while the Technology sector is over-represented (e.g. Technology Information and Media - J, Professional Services - M).

A critical gender gap emerges in several sectors (e.g. Construction-F for men, Education-P for women, Health-Q for women), reflecting the general employment trends from official data~\cite{union2020} (see Figure~\ref{fig:LinkedIn_EU_nace_gender_comparison}).
To assess these gender gaps, we compared \textsf{NACE employment GGI} (Eq.~\ref{eq:empl-NACE}) index with \textsf{LinkedIn GGI (Company industry)} (Eq.~\ref{eq:Link-attr}).
We found a strong and statistically significant correlation (Table~\ref{tab:corr_italy_linkedin_census} row 6) confirming that the gender gap in the LinkedIn data mirrors the participatory gender gap in the labour market.
Overall, diving into the distinct economic sectors, we observe that LinkedIn estimates are a reliable sensor for near real-time monitoring of the labour market dynamics in Italy. 

\subsection{Gender Gap Insights.}

\paragraph{LinkedIn Insights: Seniority.}
In Table~\ref{tab:seniority_national}, we notice that the leadership positions are not gender balanced (e.g. Manager, Owner, etc.), a finding also confirmed by Kashyap et al.~\cite{Kashyap2021} on LinkedIn estimates at a worldwide level, and social science literature on the topic~\cite{Maida2022, WEF2022}.
This may be due to the lack of economic participation and opportunity for women in apical positions~\cite{Cascella2022}.

%\textbf{RQ2} Can we enrich the current view of the gender gap in Italy as seen through LinkedIn?

\begin{figure}[!ht]
    \includegraphics[width=0.4\textwidth]{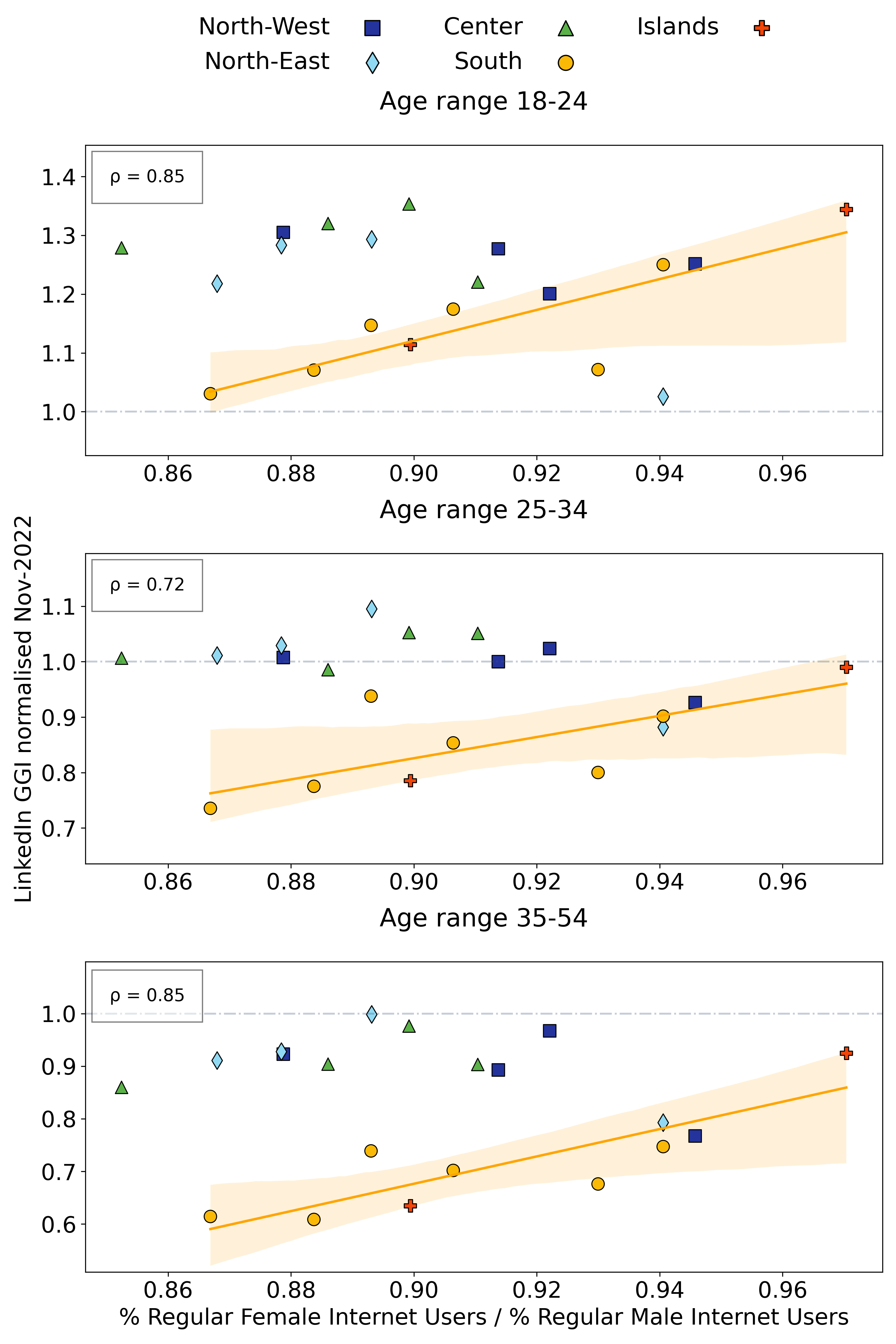}
    \caption[\textsf{LinkedIn GGI normalised} vs gender ratio of Internet users' percentages in South]{\textsf{LinkedIn GGI normalised} (y-axis) versus gender ratio F/M (x-axis) of regular Internet users' percentages by age range, region. Data were collected for Italian regions in November 2022. The yellow line shows the regression line with 95\% confidence for the regions of South and Islands, and the light blue line is the equality line.}
    \label{fig:Digitalization_corr}
\end{figure}

\paragraph{Digital Divide.}
To ensure the effectiveness of the digital divide on the obtained LinkedIn ads estimates, we employed data about the regional level of digital penetration reported by ISTAT~\footnote{Section \textit{Benessere equo e sostenibile} (BES, translation: Fair and sustainable welfare)}. 
After computing the respective ratios per gender, age, and region, we compared them to the \textsf{LinkedIn GGI normalised} (Eq.~\ref{eq:Link-GGI-Census}).
We did not observe a statistically significant correlation between the gender gap in the digitalisation data and the \textsf{LinkedIn GGI normalised} for Italy's Northern and Central regions.
On the contrary, for the Southern regions, a strong correlation between the gender gap and digitalisation level did emerge (see Table~\ref{tab:corr_italy_linkedin_census} row 4 and Figure~\ref{fig:Digitalization_corr}).
Interestingly, Fatehkia et al.\cite{Fatehkia2019}, employing data from the Facebook advertising platform, also found that the digitalisation level is a good proxy for the gender gap worldwide.

\begin{figure}[!ht]
    \includegraphics[width=0.47\textwidth]{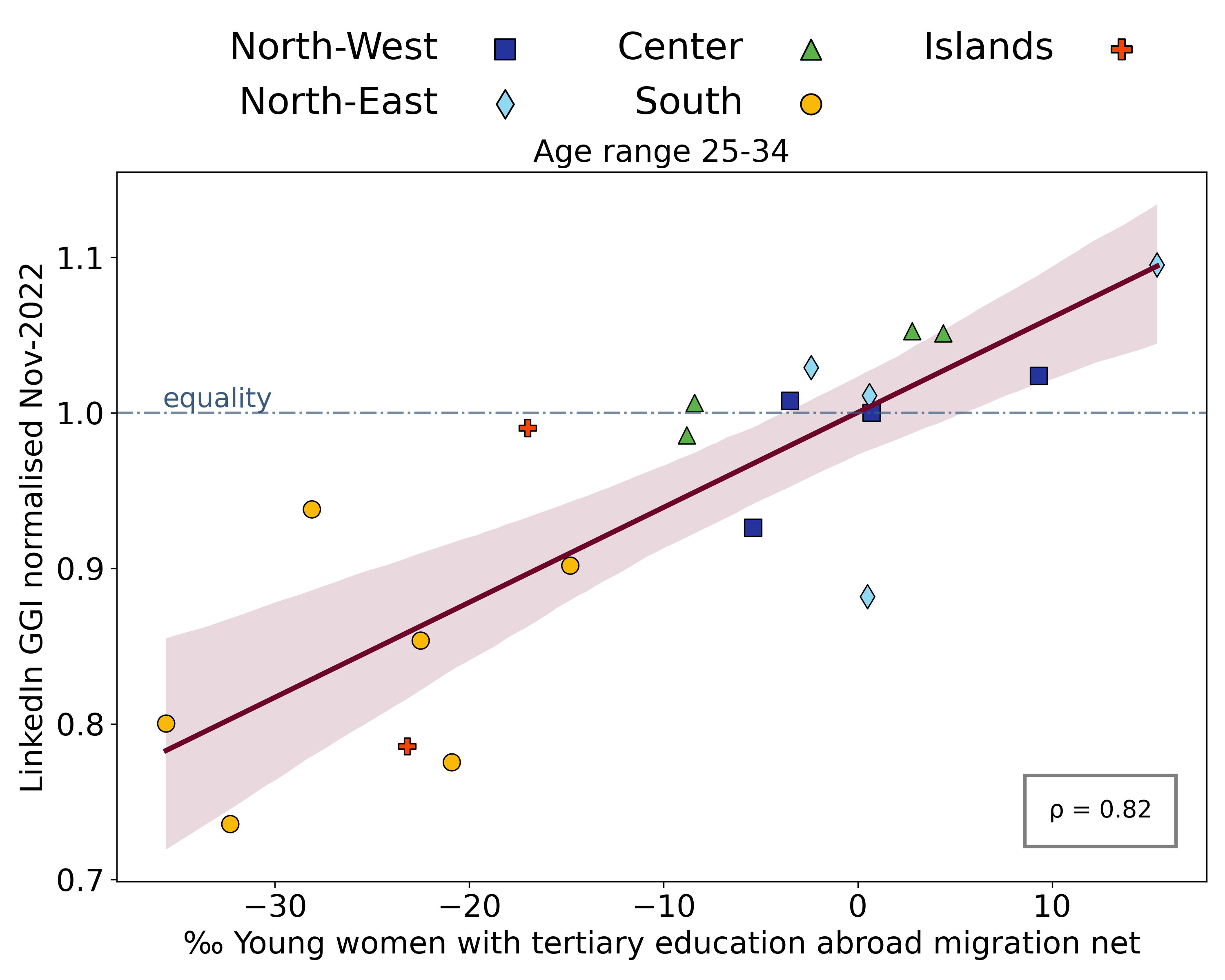}
    \caption[\textsf{LinkedIn GGI normalised} vs young female mobility]{\textsf{LinkedIn GGI normalised} (y-axis) versus young female migration net for women with tertiary education over one thousand advanced educated women from 25 to 39 years old staying in Italy. The dark red line shows the regression line for all regions of Italy with 95\% confidence, and the light blue line is the equality line. Data were collected for Italian regions in November 2022.}
    \label{fig:mobility_corr_25_34}
\end{figure}

\paragraph{Mobility.}
\label{subsec:mobility}
We obtained the mobility data for younger age groups (25 to 39 years old) with tertiary education (e.g. graduated, PhD) from the ISTAT, aggregated by gender and region. The data are provided as a migration net of highly educated individuals, indicating the difference between incoming and outgoing migration flows over one thousand individuals staying in Italy.
A negative value indicates more people moving abroad, and a positive otherwise. 

Table~\ref{tab:corr_italy_linkedin_census} (row 5), shows a significant positive correlation between the youth mobility measures of women and the \textsf{LinkedIn GGI normalised} (Eq.~\ref{eq:Link-GGI-Census}).
Figure~\ref{fig:mobility_corr_25_34} shows the relationship between the most affected age ranges (25-34) (see Appendix~\ref{sec:appendix} for the full report of age groups). In relationship with the official data, the  \textsf{LinkedIn GGI normalised} is higher where more young graduated women arrived from abroad (Centre, North regions), and lower otherwise (South, Islands).
This finding reflects a phenomenon typical for the Italian labour panorama, known as ``brain drain,'' where the highly educated youth seeks employment abroad~\cite{Cattaneo2019, Ruiu2019}. 

\subsection{Predicting Gender Gap.}
\paragraph{Predictive Modelling of the Labour Gender Gap.}\label{sec:model}
Where traditionally reported data are sparse or hard to obtain, inferring the gender gap in the labour market through the lenses of LinkedIn estimates could provide insights at scale and granularity for timely and accurate policies.
Here, we were interested in predicting the gender gap from the LinkedIn estimates not simply as population aggregates but to age group and geographic location.
We postulated the task as a multi-linear regression problem where leveraging the LinkedIn estimates; we predicted the actual gender gap.
Before training, we performed outlier detection using Isolation Forest Algorithm. 
Removing the outliers, the obtained result ($R^2 = 0.77$, $MAE = 0.05$) showed no improvement and let us deduce that the gender gap is uniform without off-scale points.
The model was trained on a dataset of four age ranges in 20 regions. The results yielded a  $R^2_{adj} = 0.75$ ($R^2 = 0.78$, $MAE = 0.05$).
The results of the 5-fold cross validation yielded an average $R^2 = 0.72$ (with standard deviation 0.03) and average $MAE = 0.05$.
Out of all data points, three residuals (difference between actual and predicted values) were more than two times greater than the residual standard error (0.06) of the distribution (Liguria 18-24; Umbria 18-24, Puglia 55+). These were all associated with the more sparse age ranges in LinkedIn (18-24 and 55+), for which the accuracy of the \textsf{LinkedIn GGI normalised} index falters.

\section{Conclusions}

Gender inequalities and discrimination in the labour market still hold despite significant advances worldwide. Given the slow pace and high cost of official surveying approaches, we assessed whether estimates from the LinkedIn advertising platform might be employed to obtain insights into the gender gaps observed in the work environment. 

Here, we focused on Italy as a case study, a country for which we have detailed census data, which is at the forefront of the European economic scene but at the same time presents substantial sociodemographic inequalities. Touching upon a series of characteristics, such as seniority level and the various industrial domains, we shed light on the strengths and limitations of such tools for understanding the sub-national labour markets.

More precisely, we showed that LinkedIn advertising audience estimates could be employed as a proxy for the Italian labour market as it reflects the official statistics concerning age and gender and demographic distribution of the labour force.
At the same time, the intrinsic biases of the platform - it is mainly adopted by highly skilled professionals and is less popular to older age groups - present several shortcomings to the industry category and seniority level, with industries such as Technology being more represented while others such as Farming to be underrepresented. 

We also highlighted essential aspects related to the gender gap in Italy; the digital divide that varies significantly within the country can be an explanatory factor of the gender gap observed in LinkedIn, as women in the Southern regions of Italy have the lowest digitalisation rate.
Last but not least, our findings underline the phenomenon of ``brain drain'', of significant concern for Italy, as the younger and skilled generation is migrating abroad. Our data show that this trend is particularly intense for the female population of the Southern areas of Italy.

\begin{acks}
    The authors gratefully acknowledge the support from the Lagrange Project of the Institute for Scientific Interchange Foundation (ISI Foundation) funded by Fondazione Cassa di Risparmio di Torino (Fondazione CRT).
\end{acks}

%\clearpage
\bibliographystyle{ACM-Reference-Format}
\bibliography{Monitoring_Gender_Gaps_via_Linkedin_Advertising_Estimates}

\appendix

\setcounter{table}{0}
\setcounter{figure}{0}
\renewcommand{\thetable}{A\arabic{table}}
\renewcommand{\thefigure}{A\arabic{figure}}

\section{Appendix}
\label{sec:appendix}

\begin{figure}[hbtp]
    \includegraphics[width=0.47\textwidth]{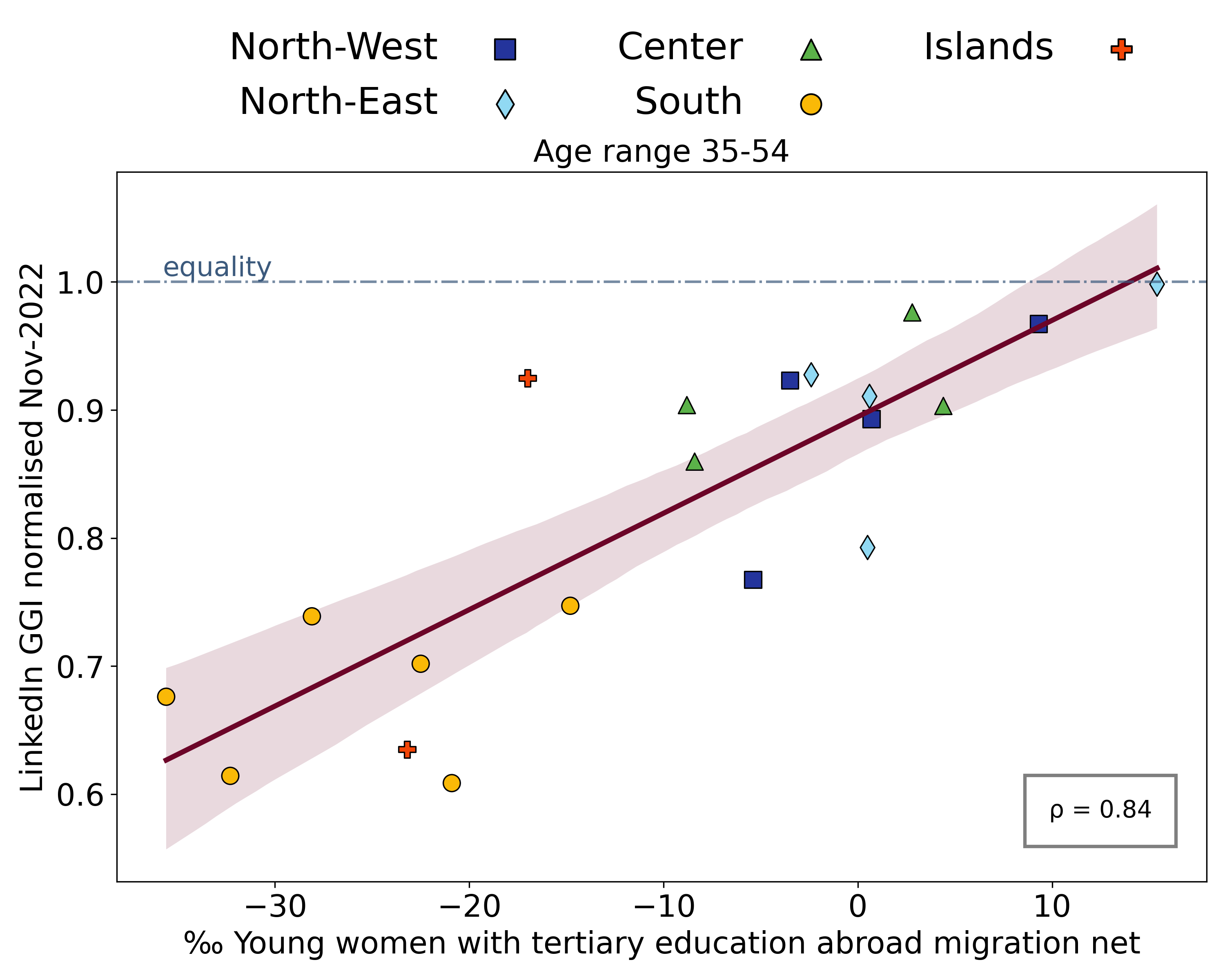}
    \caption{\textsf{LinkedIn GGI normalised} for age range 35-54 (y axis) versus young female migration net for women with tertiary education over one thousand advanced educated women from 25 to 39 years old staying in Italy. Dark red line shows regression line for all regions of Italy with 95\% confidence, light blue line is the equality line.}    \label{fig:mobility_corr_35_54}
\end{figure}

We report the choices needed to compare offline data and LinkedIn estimates. Since the age ranges were not always directly comparable among different data sources, we had to link them: the age groups match between LinkedIn and ISTAT are in Table~\ref{tab:map_age_link_ISTAT}, in Table~\ref{tab:map_age_euro_link} for LinkedIn and EUROSTAT. The \textsf{LinkedIn GGI (Seniority)} (Eq.~\ref{eq:Link-attr}) are included in Table~\ref{tab:seniority_national} by age ranges. 
Moreover, in Table~\ref{tab:map_industry_NACE} we outline the mapping of the EUROSTAT NACEs to the \emph{Company industry} attribute in LinkedIn ads platform.

Next, we incorporate all the (Pearson) aforementioned correlation values and the statistic significance associated. In addition, to evaluate the stability of the exclusion query in our data set, in Table~\ref{tab:corr_italy_linkedin_census} (row 1), we provide the correlation values for the measures computed with or without exclusion query to the level of regions. The correlation values are computed correlating the original estimates of locations that hit the 300 threshold with the values obtained applying them the exclusion query. 
Finally, concerning the correlation with mobility data exposed in Section~\ref{subsec:mobility}, the correlation among young female mobility data and LinkedIn gender gap affects also age range 35-54. Hence, we integrate Figure~\ref{fig:mobility_corr_25_34} with the plot of the regression line for age range 35-54 in Figure~\ref{fig:mobility_corr_35_54}. The Pearson's correlation values are in~\ref{tab:corr_italy_linkedin_census}, row 5.

\begin{table}[pb]
    \caption{ISTAT and LinkedIn age ranges mapped.}
    \label{tab:map_age_link_ISTAT}
    %\small
    \begin{tabular}{ l l } 
     \toprule
        \texttt{ETA1} ISTAT & LinkedIn age range\\
     \midrule
        Y15-24 & 18-24 \\
        Y25-34 & 25-34 \\
        avg(Y35-44, Y45-54) & 35-54 \\
        Y55-64 & 55+ \\
     \bottomrule
\end{tabular}
\end{table}

\begin{table}[pb]
        \caption{EUROSTAT and LinkedIn age ranges mapped.}
        \label{tab:map_age_euro_link}
        \begin{tabular}{ l l } 
         \toprule
            \texttt{AGE} EUROSTAT & LinkedIn age range\\
         \midrule
            From 15 to 24 years & 18-24 \\
            From 25 to 49 years & avg(25-34, 35-54) \\
            From 55 to 74 years & 55+\\
         \bottomrule
        \end{tabular}
\end{table}

\begin{table}[pb]
   \caption{National values of \textsf{LinkedIn GGI (Seniority)} by age range.}
   \label{tab:seniority_national}
   \small
    \begin{tabular}{ l l p{0.09\textwidth} }
    \toprule               
    Seniority & Age range  &  \textsf{LinkedIn GGI (Seniority)}\\
    \midrule
    \multirow{4}{*}{\emph{Unpaid}}
    & 18-24          & 0.50  \\
    & 25-34          & 0.36  \\
    & 35-54          & 0.43  \\
    & 55+            & 0.45  \\
    \midrule
    \multirow{3}{*}{\emph{Training}}
    & 18-24          & 1.09  \\
    & 25-34          & 1.23  \\
    & 35-54          & 1.79  \\
    \midrule
    \multirow{4}{*}{\emph{Entry}}
    & 18-24          & 1.04  \\
    & 25-34          & 1.00  \\
    & 35-54          & 1.18  \\
    & 55+            & 1.35  \\
    \midrule
    \multirow{4}{*}{\emph{Senior}}
    & 18-24          & 0.93  \\
    & 25-34          & 0.98  \\
    & 35-54          & 1.12  \\
    & 55+            & 1.01  \\
    \midrule
    \multirow{4}{*}{\emph{Manager}}
    & 18-24          & 0.64  \\
    & 25-34          & 0.57  \\
    & 35-54          & 0.71  \\
    & 55+            & 0.71  \\
    \midrule         
    \multirow{4}{*}{\emph{Director}}
    & 18-24          & 0.62  \\
    & 25-34          & 0.56  \\
    & 35-54          & 0.67  \\
    & 55+            & 0.70  \\
    \midrule
    \multirow{4}{*}{\emph{VP}}
    & 18-24          & 0.56  \\
    & 25-34          & 0.57  \\
    & 35-54          & 0.63  \\
    & 55+            & 0.61  \\
    \midrule
    \multirow{4}{*}{\emph{CxO}}       
    & 18-24          & 0.44  \\
    & 25-34          & 0.43  \\
    & 35-54          & 0.46  \\
    & 55+            & 0.54  \\
    \midrule
    \multirow{3}{*}{\emph{Partner}}
    & 25-34          & 0.65  \\
    & 35-54          & 0.68  \\
    & 55+            & 0.63  \\
    & 55+            & 0.73  \\
    \midrule 
    \multirow{4}{*}{\emph{Owner}}
    & 18-24          & 0.55  \\
    & 25-34          & 0.47  \\
    & 35-54          & 0.55  \\
    & 55+            & 0.70  \\
    \bottomrule 
\end{tabular}
\end{table}

\begin{table*}[htbp]
        \caption{Map of LinkedIn company industry, EUROSTAT NACE name and code for the corresponding ones.}
        \label{tab:map_industry_NACE}
        \small
        \begin{tabular}{p{0.33\linewidth}  p{0.53\linewidth} p{0.055\linewidth} }
            \toprule
                LinkedIn Company industry & EUROSTAT NACE & Code \\ 
            \midrule
            Farming Ranching Forestry & Agriculture, forestry and fishing & A \\ 
            Oil Gas and Mining & Mining and quarrying & B \\ 
            Manufacturing & Manufacturing & C \\ 
            Utilities & Water supply; sewerage, waste management and remediation activities & E \\ 
            Construction & Construction & F \\ 
            Wholesale & Wholesale and retail trade; repair of motor vehicles and motorcycles & G \\ 
            Retail & Wholesale and retail trade; repair of motor vehicles and motorcycles & G \\ 
            Transportation Logistics Supply Chain and Storage & Transportation and storage & H \\ 
            Accommodation & Accommodation and food service activities & I \\ 
            Technology Information and Media & Information and communication & J \\ 
            Financial Services & Financial and insurance activities & K \\ 
            Holding Companies & Other service activities & K \\  
            Real Estate and Equipment Rental Services & Real estate activities & L \\ 
            Professional Services & Professional, scientific and technical activities & M \\ 
            Administrative and Support Services & Administrative and support service activities & N \\ 
            Government Administration & Public administration and defence; compulsory social security & O \\ 
            Education & Education & P \\ 
            Hospitals and Health Care & Human health and social work activities & Q \\ 
            Entertainment Providers & Arts, entertainment and recreation & R \\
             \bottomrule
        \end{tabular}
\end{table*}

\begin{table*}[htbp]
        \caption{Correlation values between LinkedIn ads data and offline data. 
    Notation for the significance is the following: *** p-value $< 0.001$, ** p-value $< 0.01$, * p-value $< 0.05$}
        \label{tab:corr_italy_linkedin_census}
        \begin{tabular}{p{0.5\textwidth} l l l }
         \toprule
        Indices & Age Range & Pearson coefficient $\rho$ \\
        %\multirow{4}{=}{\emph{(1)} \emph{\textsf{LinkedIn GGI} with exclusion query versus \textsf{LinkedIn GGI} without exclusion query NUTS3}}
        %& 18-24  & 0.79 *** \\
        %& 25-34  & 0.97 *** \\
        %& 35-54  & 0.93 *** \\
        %& 55+    & 0.99 *** \\
        \midrule 
        \multirow{4}{=}{\emph{(1)} \emph{\textsf{LinkedIn GGI normalised} with exclusion query versus \textsf{LinkedIn GGI normalised} without exclusion query NUTS2}}
        & 18-24  & 0.92 *** \\
        & 25-34  & 0.97 *** \\
        & 35-54  & 0.98 *** \\
        & 55+    & 0.99 *** \\
        \midrule 
        %\multirow{4}{=}{\emph{(3)} \emph{\textsf{LinkedIn GGI normalised} vs ISTAT Census employment NUTS3}} 
        % & 18-24  & 0.33 *** \\
        % & 25-34  & 0.71 *** \\
        % & 35-54  & 0.82 *** \\
        % & 55+    & 0.67 *** \\      
        \multirow{4}{=}{\emph{(2)} \emph{\textsf{LinkedIn GGI normalised} vs \textsf{Employment GGI}}} 
        & 18-24  & 0.57 ** \\
        & 25-34  & 0.81 *** \\
        & 35-54  & 0.83 *** \\
        & 55+    & 0.63 ** \\
        \midrule         
        \multirow{3}{=}{\emph{(3)} \emph{\textsf{\% LinkedIn members working in Company industry by gender} vs EUROSTAT \% employees working in NACE by gender}}
        & 18-24 & 0.67 *** \\
        & 25-49 & 0.70 *** \\
        & 55+   & 0.48 ** \\
        \midrule 
        \multirow{3}{=}{\emph{(4)} \emph{\textsf{LinkedIn GGI normalised} vs Gender Ratio Regular Internet Users in South NUTS2}} 
        & 18-24  & 0.85 ** \\
        & 25-34  & 0.72 * \\
        & 35-54  & 0.85 ** \\
        \midrule 
        \multirow{2}{=}{\emph{(5)} \emph{\textsf{LinkedIn GGI normalised} vs Female youth mobility (NUTS2)}}  
        & 25-34  & 0.82 *** \\
        & 35-54  & 0.84 *** \\  
        \midrule
        \multirow{3}{=}{\emph{(6)} \emph{\textsf{LinkedIn GGI (Company industry)} vs \textsf{NACE Employment GGI}}}
        & 18-24 & 0.88 *** \\
        & 25-49 & 0.91 *** \\
        & 55+   & 0.83 ** \\
        \bottomrule 
        \end{tabular}
\end{table*}

\end{document}